\documentclass[aps,prev,twocolumn,preprintnumbers,floatfix,nofootinbib]{revtex4-1}
\pdfoutput=1
\usepackage{graphicx}
\usepackage{bm}
\usepackage{times}
\usepackage{slashed}
\usepackage{color}
\usepackage{slashed}
\usepackage{amsmath}
\usepackage{amsthm}
\usepackage{subfigure}

\newcommand{\crn}{\nonumber \\}

\newcommand{\al}{\alpha}
\newcommand{\la}{\lambda}

\newcommand{\om}{\omega}
\newcommand{\pa}{\partial}
\newcommand{\fr}{\frac}

\newcommand{\be}{\begin{equation}}
\newcommand{\ee}{\end{equation}}
\newcommand{\bea}{\begin{eqnarray}}
\newcommand{\eea}{\end{eqnarray}}

\newcommand{\ep}{\epsilon}

\begin{document}

\title{Muon Magnetic Moment and Lepton Flavor Violation in the Economical 3-3-1 Model}

\author{D. Cogollo$^{a}$}
\email{diegocogollo@df.ufcg.edu.br}

\affiliation{$^a$Departamento de F\'isica, Universidade Federal de Campina Grande, Caixa Postal 10071,
58109-970, Campina Grande, PB, Brazil\\}

\begin{abstract}
In this work we compute all relevant contributions stemming from the economical 3-3-1 model to the muon magnetic moment and the lepton flavor violation decay $\mu \rightarrow e\gamma$. Using the current bounds on these phenomena, we derive lower limits on the scale of symmetry breaking of the model. Moreover, taking into account existing limits from meson and collider studies we show that there is still room for a possible signal in $\mu \rightarrow e\gamma$ in the near future.
\end{abstract}


\maketitle

\section{Introduction}
\label{intro}

The Standard Model (SM) has passed all precision tests thus far and therefore it provides an accurate descriptions of the fundamental laws of nature. Although, we have observational and experimental evidences for going beyond the standard model such as the existence of neutrino masses \cite{Fukuda:1998mi,Ahmad:2002jz} and dark matter \cite{Sofue:2000jx,Queiroz:2016sxf}. From that perspective 3-3-1 models are quite plausible extensions of the SM. 3-3-1 models stand for electroweak extensions of the SM, where left-handed fermions are arranged in the fundamental representation of $SU(3)_{L}$. Such models can naturally explain the number of generations \cite{Pisano:1991ee,Foot:1992rh}, might offer plausible dark matter candidates with gripping phenomenology \cite{Mizukoshi:2010ky,Huong:2011xd,Alvares:2012qv,Hooper:2012sr,Profumo:2013sca,Queiroz:2013zva,Queiroz:2013lca,Alves:2013tqa,Dong:2014wsa,Gonzalez-Morales:2014eaa,Queiroz:2014pra,Cogollo:2014jia,Alves:2015pea,Baring:2015sza,Mambrini:2015sia,Alves:2015mua,Allanach:2015gkd,Klasen:2016qux,Altmannshofer:2016jzy,Queiroz:2016sxf,Queiroz:2016awc,Profumo:2016idl,Queiroz:2016zwd,Alves:2016cqf,Arcadi:2017kky,Campos:2017odj,Arcadi:2017atc,Arcadi:2017kky}, explain neutrino masses through the seesaw mechanism \cite{Schechter:1980gr,Queiroz:2010rj}, feature interesting connections to cosmology \cite{Hooper:2011aj,Kelso:2013nwa,Kelso:2013paa,Queiroz:2014ara,Allahverdi:2014bva} and prospects to collider physics \cite{Alves:2011kc,Cogollo:2012ek,Alves:2012yp,Queiroz:2016gif,Caetano:2013nya}, among others \cite{Dong:2007qc,Dong:2007ba,VanSoa:2008bm,Dong:2011pn,Dong:2013wca,Cogollo:2013mga,Queiroz:2014yna,Dong:2014esa,Queiroz:2014ara,Phong:2014ofa,Montero:2014uya,Alves:2014yha,Long:2015qza,Dong:2015jxa,Hernandez:2016eod}. In this work we will focus our attention on the ecoomical 3-3-1 model, which has in its scalar sector two scalar triplets \cite{Dong:2012bf}, and discuss the long standing discrepancy on the muon anomalous magnetic moment and the lepton flavor violating decay $\mu \rightarrow e \gamma$ (see \cite{Lindner:2016bgg} for a recent review).

The muon anomalous magnetic moment, g-2, is one of the most precisely measured quantities in particle physics, reaching a precision of 0.54 ppm. Ever since the first experimental limits were reported, a discrepancy between the SM prediction and the experimental value on g-2 has been observed. Today this anomaly is in the ballpark of $3.6\sigma$, leading to several speculations in the context of 3-3-1 models \cite{Ky:2000ku,Kelso:2013zfa,Dong:2014wsa,Kelso:2014qka,Queiroz:2014zfa,DeConto:2016ith}. In this work we will assess the possibility of explaining g-2 in the context of the economical 3-3-1 model. 

A much more appealing phenomena is lepton flavor violation. The existence of lepton flavor violation (LFV) has tremendous implications to particle physics, since any signal of LFV would constitute an irrefutable proof the existence of new physics not far from the TeV scale. Given the current limits on LFV, new physics effects should live at energies above the TeV scale. In this work we focus the attention on the $\mu \rightarrow e \gamma$ rate since it has a much larger rate than other LFV observables \cite{Lindner:2016bgg}. Our goal as far as LFV is concerned is to derive limits on the scale of symmetry breaking of the economical 3-3-1 model and check whether a possible signal in this decay mode can be originated in this model in the light of current and future constraints from other sources. 

The paper is structured as follows: In section \ref{model} we discuss the model; in section \ref{bounds} we address existing constraints and discuss future experimental sensitivities. In section \ref{muong2} we present our results for g-2; in section \ref{decaymey} we discuss LFV; finally in section \ref{conclu} we draw our conclusions.

\section{The Economical 3-3-1 Model}
\label{model}
The Economical 3-3-1 model is a model that inherits the fermion content of the well-known 3-3-1 model with right-handed neutrinos but featuring a reduced scalar sector, thus anomaly free. In what follows we discuss separately the key ingredients that will allow the reader to follow our reasoning.

\subsection{Fermion Fields}
 The model has the following particle content,

\bea \psi_{iL}&=&\left(%
\begin{array}{c}
  \nu_i \\
  e_i \\
  \nu^c_i \\
\end{array}%
\right)_L\sim \left(3,-\frac{1}{3}\right), \mbox{ }e_{iR}\sim
(1,-1),\mbox{ }
\crn Q_{1L}&=&\left(%
\begin{array}{c}
  u_1 \\
  d_1 \\
  U \\
\end{array}%
\right)_L\sim \left(3,\frac{1}{3}\right),\mbox{ } Q_{\al L}=\left(%
\begin{array}{c}
  d_\al\\
  -u_\al\\
  D_\al\\
\end{array}%
\right)_L\sim (3^*,0),\mbox{ } \crn u_{i
R}&\sim&\left(1,\frac{2}{3}\right), \mbox{ }d_{i R} \sim \left(1,-\frac{1}{3}\right)\crn && U_{R}\sim \left(1,\frac{2}{3}\right),\mbox{
}D_{\al R} \sim \left(1,-\frac{1}{3}\right).\eea 
where $i=1,2,3$, and $\al=2,3.$. The values in
the parentheses  represent the quantum numbers under the
$\left(\mbox{SU}(3)_L,\mbox{U}(1)_X\right)$ symmetry. In this
model the electric charge operator takes a form,
\be
Q=T_3-\fr{1}{\sqrt{3}}T_8+X,\label{eco}\ee 
where $T_a$ $(a=1,2,...,8)$ are the generators of $SU(3)$ and $X$ the charged under $\mbox{U}(1)_X$.  The exotic quarks $U$ and $D_\al$ have the same electric charge as usual up and down quarks, i.e. with $q_{U}=2/3$ and $q_{D_\al}=-1/3$.

\subsection{Scalar Sector}

The scalar sector of the model is comprised of two scalar triplets. The pattern of spontaneous symmetry breaking is via two steps. 
Firstly, the scalar triplet \be \chi=\left(%
\begin{array}{c}
  \chi^0_1 \\
  \chi^-_2 \\
  \chi^0_3 \\
\end{array}%
\right) \sim \left(3,-\fr 1 3\right)\ee developing the non-trivial $vevs$
\be
\langle\chi\rangle=\fr{1}{\sqrt{2}}\left(%
\begin{array}{c}
  u \\
  0 \\
  \om \\
\end{array}%
\right),\label{vevc}\ee and in a second step the scalar triplet \be
\phi=\left(%
\begin{array}{c}
  \phi^+_1 \\
  \phi^0_2 \\
  \phi^+_3 \\
\end{array}%
\right)\sim \left(3,\fr 2 3\right)\ee develops a $vev$ as follows,
\be \langle\phi\rangle=\fr{1}{\sqrt{2}}\left(%
\begin{array}{c}
  0 \\
  v \\
  0 \\
\end{array}%
\right).\label{vevp}\ee 

These scalars form the scalar potential,

\bea V(\chi,\phi) &=& \mu_1^2 \chi^\dag \chi + \mu_2^2
\phi^\dag \phi + \la_1 ( \chi^\dag \chi)^2 + \la_2 ( \phi^\dag
\phi)^2\crn &  & + \la_3 ( \chi^\dag \chi)( \phi^\dag \phi) +
\la_4 ( \chi^\dag \phi)( \phi^\dag \chi). \label{poten} \eea

Notice that just two scalar triplets simplifies greatly the scalar potential. For this reason the economical model is a truly attractive 3-3-1 model.

\subsection{Fermion Masses}

With the scalar sector above the fermion gain masses through the Yukawa lagrangian below, 

\bea {\cal L}_Y &=&h'_{11}\overline{Q}_{1L}\chi
U_{R}+h'_{\al\beta}\overline{Q}_{\al L}\chi^* D_{\beta R}\crn
&&+h^e_{ij}\overline{\psi}_{iL}\phi
e_{jR}+h^\ep_{ij}\ep_{pmn}(\overline{\psi}^c_{iL})_p(\psi_{jL})_m(\phi)_n\crn
&&+h^d_{1i}\overline{Q}_{1 L}\phi d_{i R}+h^d_{\al
i}\overline{Q}_{\al L}\phi^* u_{iR},\crn
&&+ h^u_{1i}\overline{Q}_{1L}\chi u_{iR}+h^u_{\al i}\overline{Q}_{\al L}\chi^* d_{i
R} \crn
&&+h''_{1\al}\overline{Q}_{1L}\phi D_{\al R}+h''_{\al
1}\overline{Q}_{\al L}\phi^* U_{R}+h.c.\label{y2}\eea 

Notice that the exotic quarks $U$ and $D_\al$ have masses proportional to the $vev$ $\om$, whereas the SM fermions to $u,v$. Therefore, $\om \gg u,v$.  One may realize that the Yukawa lagrangian features a global symmetry ($L^\prime$) which is related to the lepton number (L) through,

\be
L^\prime=L - \fr{4}{\sqrt{3}}T_8.
\label{Eq:L}\ee

Using Eq.\ref{Eq:L} one finds,

\bea && L^\prime (\psi_{iL},Q_{1L}, Q_{\al
L},\phi,\chi,e_{iR},u_{iR},d_{iR},U_R,D_{\al R})=\nonumber\\
&&\fr 1 3,-\fr 2
3,\fr 2 3,-\fr 2 3,\fr 4 3,1,0,0,-2,2.\eea

The field $\chi_1^0$ carries two units of lepton number, thus a bilepton. Since this global symmetry is broken by the $vev$ $u$, then $u$ is a sort of lepton-number violating parameter, which should be very small. In our procedure we take $v = 246$~GeV.

Anyways, it has been shown that this Yukawa lagrangian sucessfully explain the fermion masses according to data \cite{Dong:2006gx,Dong:2006mt}

\subsection{Gauge Bosons}

The covariant derivative of the scalar triplets is given by \bea D_\mu &=&\pa_\mu-igT_aW_{a\mu}-ig_X T_9 X B_\mu \eea here the gauge fields $W_a$ and $B$ transform as the adjoint representations of $\mathrm{SU}(3)_L$ and $\mathrm{U}(1)_X$, with the corresponding gauge coupling constants $g$, $g_X$. Having in mind that $T_9=\fr{1}{\sqrt{6}}\mathrm{diag}(1,1,1)$, expanding this covariant derivative we get,

\bea
\fr{g}{2}\left(%
\begin{array}{ccc}
  A
  & \sqrt{2} W^+_\mu & \sqrt{2}X'^0_\mu \\
  \sqrt{2}W^-_\mu & B & \sqrt{2}W'^-_\mu \\
  \sqrt{2}X'^{0*}_\mu & \sqrt{2}W'^+_\mu &
  C \\
\end{array}%
\right),\eea where $t\equiv g_X/g$, $A\equiv W_{3\mu}+\fr{1}{\sqrt{3}}W_{8\mu}+t\sqrt{\fr 2 3}XB_\mu,B\equiv -W_{3\mu}+\fr{1}{\sqrt{3}}W_{8\mu}+
  t\sqrt{\fr 2 3}X B_\mu$, $C\equiv -\fr{2}{\sqrt{3}}W_{8\mu}+t\sqrt{\fr 2 3}X B_\mu$, and 
  
  \bea W^{\pm} _\mu &\equiv& \fr{W_{1\mu}\mp
iW_{2\mu}}{\sqrt{2}},\crn W'^\mp_\mu &\equiv& \fr{W_{6\mu}\mp
iW_{7\mu}}{\sqrt{2}}, \crn X'^0_\mu &\equiv&
\fr{W_{4\mu}-iW_{5\mu}}{\sqrt{2}}. \eea

Since we are investigating an $SU(3)\otimes U(1)_X$ gauge group there are in total nine gauge bosons with four of them belonging to the SM spectrum ($W^\pm,Z,A$). The new gauge bosons are the heavy charged gauge boson $W^{\prime\pm}$, the electrically neutral $X^0$ that carries two units of lepton number, and a heavy $Z^\prime$ boson.

In the limit $\omega \gg u,v$ the masses of the gauge bosons are easily obtained and read,

\bea M^2_{W}&=&\fr{g^2v^2}{4},\\
M^2_{W^\prime}&=&\fr{g^2}{4}(u^2+v^2+\om^2),\\
M^2_{X^0}&=&\fr{g^2}{4}(\om^2+u^2),\\
M_{Z^{\prime}}^2 &=& \frac{g^2 c^2_w w^2}{3-4s_w^2}.
\label{massy}
\eea

This is the gauge boson spectrum of the model and these gauge bosons are the main characters of our phenomenology. Before discussing g-2 and LFV we need to present the neutral and charged currents.

\subsection{Neutral and Charged Currents}

The neutral and charged currents arise from the kinect terms of the fermions, $\bar{\psi_L}D_\mu \gamma^\mu \psi_L + \bar{\psi_r}D_\mu \gamma^\mu \psi_R$, yielding,

\be {\cal L}^{NC} \supset
\bar{f}\, \gamma^{\mu} [g_{Vf} + g_{Af}\gamma_5]\, f\,
Z'_{\mu}. \label{ncm} \ee with, 
\be
g_{Vf}= \frac{g}{4 c_W} \frac{(1 -
4 s_W^2)}{\sqrt{3-4s_W^2}},\
g_{Af} = -\frac{g}{4 c_W \sqrt{3-4s_W^2}},
\label{hsr}
\ee and

\be
{\cal L}^{CC}_l \supset - \frac{g}{2\sqrt{2}}\left[
\bar{\nu^c}\gamma^\mu (1- \gamma_5)\, l \, W^{\prime -}_\mu
 +h.c. \right],
\label{Wprime}
\ee 

Obvisouly Eq.\ref{ncm} and Eq.\ref{Wprime} are not the complete current of the model. There are more terms involving quarks, and neutrinos but these will not be relevant for our discussion which is concentrated on the charged leptonic sector. 

We have gather all important ingridient for our g-2 and LFV computation. Thus we now move these phenomena.

\section{Existing Bounds}
\label{bounds}
\subsection{Meson Decays}
With the enormous improvement over the experimental precision on meson decays, new physics contributions to rare meson decays can now be tested. In particular, data on the meson B decays $B_{s,d} \rightarrow \mu^+ \mu^-$ and $B_{d} \rightarrow K^{\star} (K)\mu^+ \mu^-$ turned out to be great laboratories to test the existence of new vector gauge bosons \cite{Buras:2014yna}.  In light of no significant deviation over the SM predictions, bounds were derived on the $Z^\prime$ mass, excluding $Z^{\prime}$ below $\sim 2-3$~TeV. The uncertainty in the bound stems from the depedence on the parametrization in the quark mixing matrices \cite{Buras:2014yna}.  

\subsection{Dilepton}

In addition to meson physics, the advent of the Large Hadron Collider (LHC) has set a new era in the search for new physics \cite{Lindner:2016lpp}. In particular, both ATLAS and CMS collaborations have searched for neutral vector gauge bosons in the dilepton channel ($ee$ and $\mu\mu$) finding no evidence, setting stringent lower mass bounds on the $Z^\prime$ mass \cite{Patra:2015bga,Khachatryan:2016zqb,ATLAS:2017wce} of various models. A speficic study for the economical 3-3-1 model was peformed in \cite{Queiroz:2016gif,Alves:2016fqe,Cao:2016uur}. There the authors found $M_{Z^\prime} > 3.8$~TeV for $13fb^{-1}$ of integrated luminosity, possibly reaching $M_{Z^\prime} > 4.9$~TeV and $M_{Z^\prime} > 6.1$~TeV for $100^{-1}$ and $1000fb^{-1}$ integrated luminosity, respectively.

\subsection{Charged Lepton + MET}

The economical 3-3-1 model predicts the existence of a charged gauge boson that interacts with charged leptons as shown in Eq.20. Such a $W^\prime$ when produced at the resonance decays into a charged lepton plus a neutrino. Therefore, searches for charged lepton + MET are suitable for these charged gauge bosons \cite{Queiroz:2016qmc,Lindner:2016lxq}. Since no excess has been observed,  a lower mass bound of $4.76$~TeV was found  with $13.3\,fb^{-1}$ of luminosity at $13$~TeV of center of mass energy \cite{Alves:2016fqe}. Moreover, future limits were projected for $100$ and $1000\, fb^{-1}$ of data, which would lead the exclusion of $W^\prime$ masses below $5.8$~TeV and $7$~TeV \cite{Alves:2016fqe}.

Now we outlined the most stringent limits on the mass of the gauge bosons we will discuss the g-2 and $\mu \rightarrow e\gamma$ decay.

\begin{figure}[!t]
\centering
\includegraphics[width=\columnwidth]{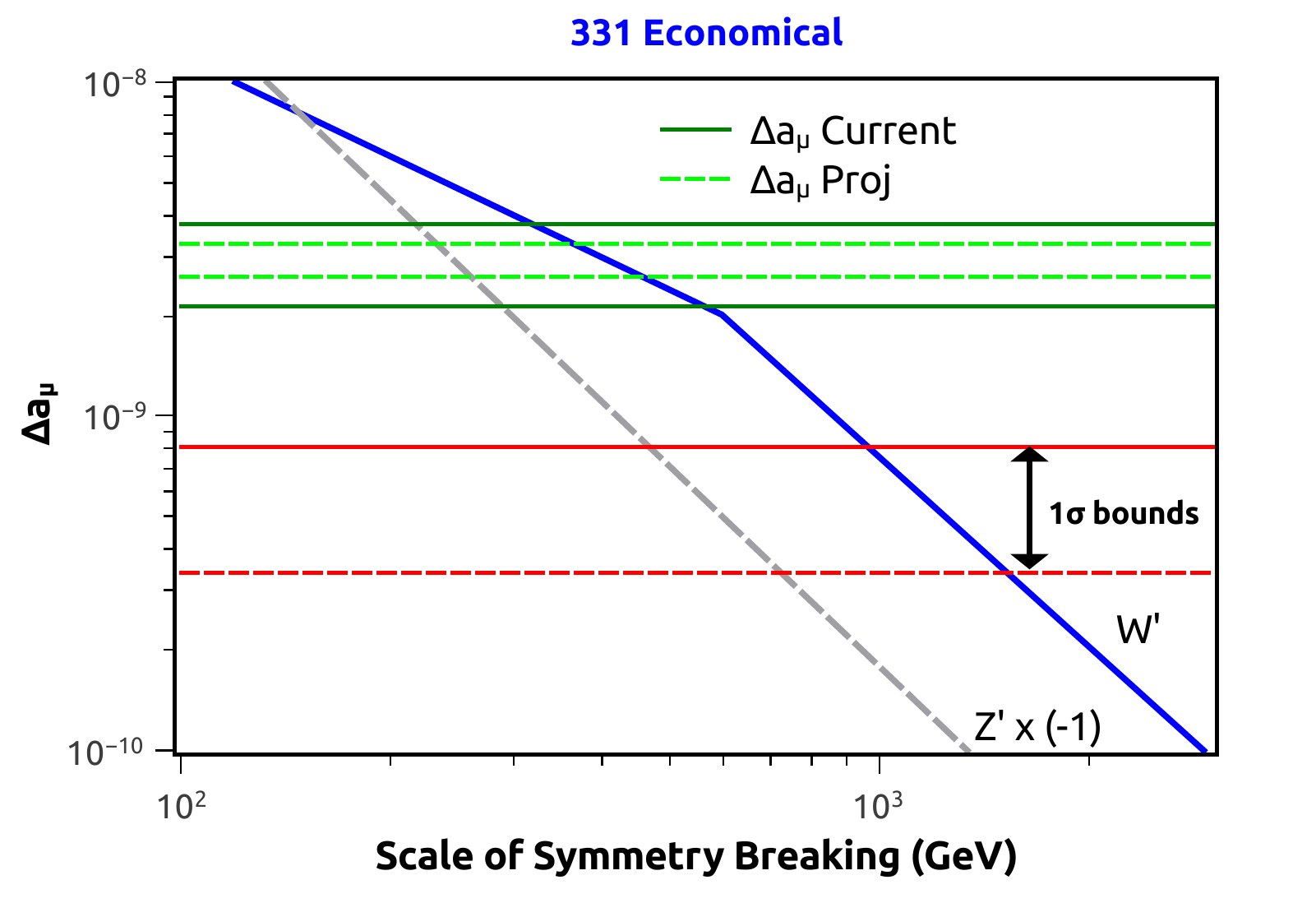}
\caption{Individual contributions from the 331 Economical model as a function of the scale of symmetry breaking. The $Z^\prime$  and $W^\prime$ contributions are negative and positive, respectively. Since the $W^\prime$ correction to g-2 is larger, the overall correction to $g-2$\, is positive.} \label{Graph4}
\end{figure}

\section{Muon Anomalous Magnetic Moment}
\label{muong2}
Fundamental charged particles feature a magnetic dipole moment ($g$) which according to classical quantum mechanics should be equal two. However, in the framework of relativistic quantum mechanism there are quantum corrections beyond the tree-level which deviates g from two. This deviation is parametrized in terms of $a = (g-2)/2$, known as the anomalous magnetic moment, for short $g-2$.

Interestingly the SM prediction for g-2 ($a_{\mu SM}$) does not agree with the experimental measurement ($a_{\mu exp}$) at the $3.6\sigma$ level \cite{Blum:2013xva}  pointing to,

\begin{equation}
\Delta a_\mu= a_{\mu exp}- a_{\mu SM} = (287 \pm 80)\times 10^{-11}.
\end{equation}

This long standing discrepancy has trigged several model building efforts. Fortunately, the ongoing $g-2$ experiment at FERMILAB will shed light into this
problem in the upcoming years. If the central value remains intact, a $5\sigma$ evidence for new physics would result, with
$\Delta a_{\mu} =(287 \pm 34)\times 10^{-11}$. Thus it is worthwhile to assess which models this discrepancy can address this anomaly in agreement with current and planned experimental limits.

In our model, the main particles contributing to g-2 are the neutral and electrically charged gauge bosons. The scalar particles in the model lead to very suppressed contributions to g-2 because their couplings to the muon are proportional to the muon mass. That said, the gauge boson corrections to $g-2$ are found to be \cite{Lindner:2016bgg},

\begin{equation}
\Delta a_{\mu}(Z^{\prime}) = \frac{m_{\mu}^2}{4 \pi^2 M_Z^{\prime 2}}\left(\frac{1}{3}g^2_{V\mu} - \frac{5}{3}g^2_{A\mu}\right).
\label{vectormuon3}
\end{equation}

\begin{eqnarray}
&&
\Delta a_{\mu} (W^{\prime}) = \frac{1}{4\pi^2}\frac{m_\mu^2}{ M_{W^{\prime}}^2 } \left[g_{V}^2 \left( \frac{5}{6} - \frac{m_{\nu}}{m_{\mu}}\right)  + g_{A}^2 \left(  \frac{5}{6} + \frac{m_{\nu}}{m_{\mu}} \right) \right],\nonumber\\
\label{vectormuon6}
\end{eqnarray} where $g_V$ and $g_A$ are the vector and vector-axial couplings in Eq. \eqref{hsr} and Eq. \eqref{Wprime}. 

With these equations at hand we compute the contributions of these gauge bosons to $\Delta a_\mu$. These contributions are shown in Fig.\ref{Graph4} as function of the scale of symmetry breaking of the model. Moreover, we overlay the current (projected) $1\sigma$ error band, that reads $80\times10^{-11} (34 \times 10^{-11})$, to derive bounds on the scale of symmetry based upon the assumption that the anomaly is otherwise resolved by any other means.

From Fig.\ref{Graph4} we see that $Z^\prime$ $(W^\prime)$ give rise to a negative (positive) correction to g-2. However, since the contribution from $W^\prime$ 331 Economical model is larger, it generates an overall positive contribution to g-2. The scale of symmetry below 1TeV needed to accommodate the anomaly is too small and it hass been excluded by other data sets. Therefore, the 331 Economical model cannot trully accommodate $g-2$. To clearly note this statement lets take a closer look into Eq.17. From this equation we get that $M_Z^\prime \approx 0.4$ $w$. Hence the limit of 3.8TeV on the $Z^\prime$ implies $w \geq 9$TeV, making it impossible to accommodate g-2 in the economical 331 model, since we needed $w < 1$TeV to do so. Moreover, the bound we get on $w$ by imposing the $1\sigma$ error bar aforementioned lead to a lower limit of $w > 1.4$~TeV which is less competetive than collider searches.

\section{$\mu \rightarrow e\gamma$ decay}
\label{decaymey}

In the SM the lepton flavor is a conserved quantity and neutrinos are massless. Although, neutrinos experience flavor oscillations
\cite{Forero:2014bxa,Kajita:2016cak,McDonald:2016ixn} constituting an experimental confirmation that lepton flavor is violated. The mechanism responsible for lepton flavor violation is completely unknown but there are some proposal in the literature \cite{Pascoli:2003uh,Petcov:2005jh,Lindner:2016lxq}. Anyways, an observation of charged LFV would necessarily imply into new physics with huge implications to model building endeavours.  

The charged current mediated by the $W^\prime$ gauge boson might induce the non-observed decay $\mu \to e\gamma$. The non-observation of this decay yields tight bounds on new physics effects. Indeeed, current (projected) bound from MEG collaboration reads \cite{TheMEG:2016wtm},
${\rm Br(\mu \to e\gamma)} < 4.2 \times 10^{-13}\ (4 \times
10^{-14})$. Adapting the results from \cite{Lindner:2016bgg} to our model we get, 

\begin{eqnarray}
{\rm Br(\mu \to  e\gamma)} = 6.43\times 10^{-6}\left(\frac{1\ \mathrm{TeV}}{M_{W'}}\right)^4\sum\limits_{f}(g^{fe*}g^{f\mu})^2,\nonumber\\
\end{eqnarray}with $g^{fe}=g\,U^{Ne\ast}/(2\sqrt{2})$ and $g^{f\mu}=g\,U^{N\mu\ast}/(2\sqrt{2})$.\\[-.2cm]

\begin{figure}[h]
 \centering
\includegraphics[width=\columnwidth]{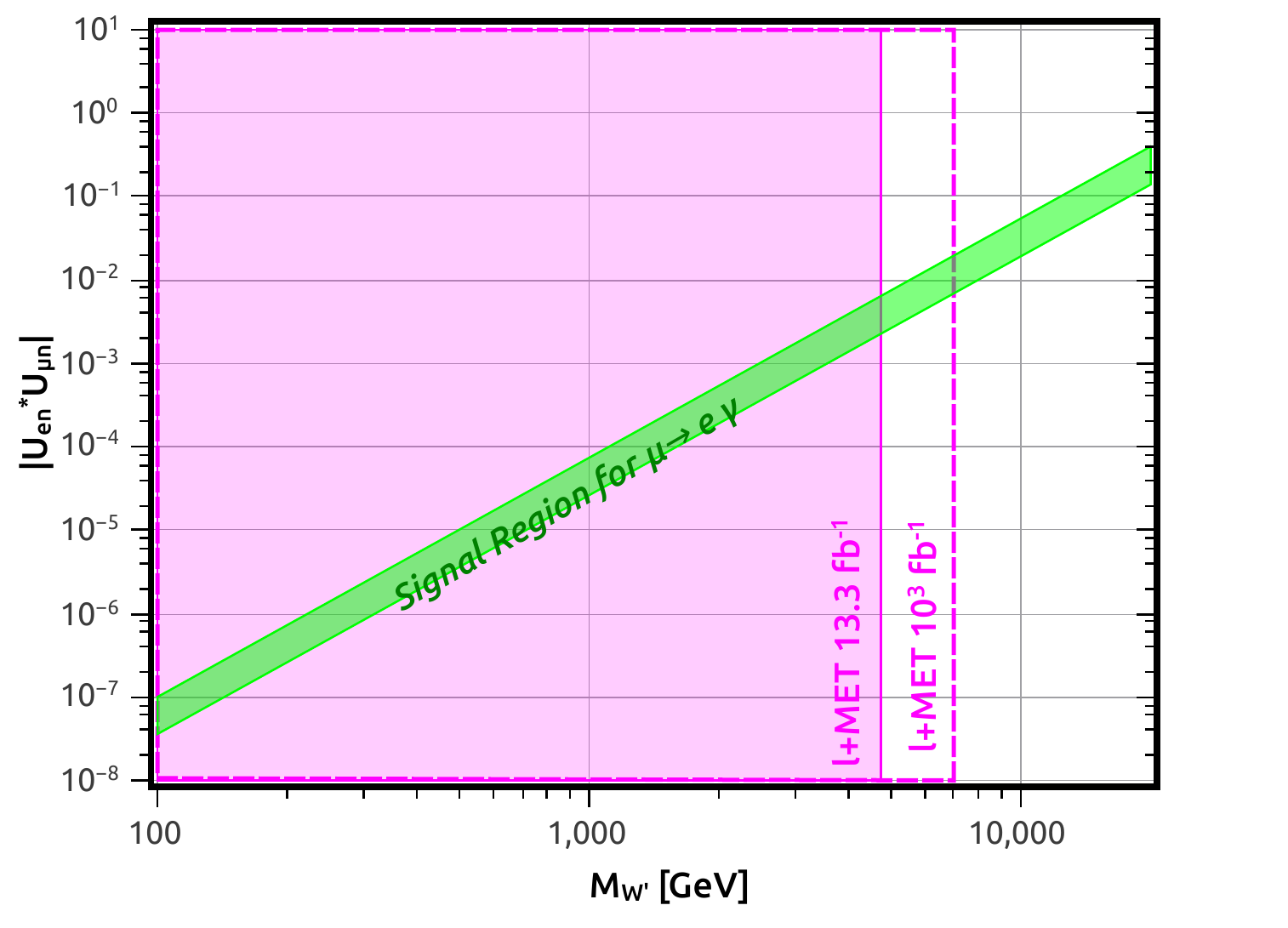}
 \caption{Region of parameter space with
   $4.2 \times 10^{-13} < {\rm Br(\mu \to e\gamma)} < 4 \times
   10^{-14}$ in green. The current (projected) bound from l+MET searches at the LHC  is shown as shaded region (dashed line).}
 \label{fig2}
\end{figure}

Hence, one can use the experimental bound on this branching ratio to place a restrictive limit on the product
$U^{Ne\ast}U^{N\mu}$ as function of the $W^{\prime}$ mass as shown in
Fig.\ref{fig2}. There, we overlaid the current collider limits on the $W^\prime$ mass for $13.3fb^{-1}$ and the projected for $1000 fb^{-1}$ as described before, as well as the possible signal region for the $\mu \rightarrow e \gamma$ decay which is delimited by the current and future sensitivity values for $Br(\mu \to  e\gamma)$.

From Fig.\ref{fig2} we can conclude that:

(i) Depending on value of the product $U^{Ne\ast}U^{N\mu}$ of interest, the $\mu \to e\gamma$ observable can outperform collider probes in the search for $W^\prime$ gauge bosons. 

(ii) The observation of possible signal in $\mu \to e\gamma$ decay might be accommodate within the economical 3-3-1 model in agreement with current and foreseen limits.

\section{Conclusions}
\label{conclu}
The economical 3-3-1 model is an attractive model where the number of generations is addressed while featuring a reduced scalar spectrum in comparison with other incarnations of 3-3-1 models. In this work we computed the relevant contributions to the muon anomalous magnetic moment to show that the Economical 3-3-1 model, while generating a positive contribution to g-2, it  cannot accommodate the anomaly since the scale of symmetry breaking needed to explain g-2 has been excluded by LHC probes for new physics. 

Moreover, we have investigated the $\mu \rightarrow e \gamma$ decay to conclude that $\mu \rightarrow e \gamma$ observable provides an interesting probe for new physics, particularly complementary to collider searches. Lastly, we found that in case of a positive signature of $\mu \rightarrow e \gamma$  in the foreseeable future, the Economical 3-3-1 model offers a plausible new physics interpretation in agreement with current and future experimental limits.

\section*{Acknowledgement}

The author would like to thank Farinaldo Queiroz for useful discussions.

\bibliographystyle{ieeetr}
\bibliography{combined2}

\end{document}